\documentclass{elsart}
\usepackage{psfig}
\usepackage{natbib}

\newcommand{\hbeta}{H$_{\beta}$}
\newcommand{\kms}{km\,s$^{-1}$}
\newcommand{\MB}{M$_{\rm B}$}

\begin{document}
\begin{frontmatter}

\title{BLR  velocities in optically and X-ray selected AGN samples}

\author[PSU]{Dieter Engels and Ralf Keil}
\address[PSU]{Hamburger Sternwarte, Gojenbergsweg 112, D-21029 Hamburg, Germany}

\begin{abstract}
We have analyzed optical spectra of 473 X-ray and 235 optically
selected AGNs, to study their emission line properties. We present 
results of an  analysis of the \hbeta\ linewidths. We find 
that the linewidth distribution of quasars is shifted towards higher
velocities ($<$v$>$=4300 \kms) compared to the distribution of Sy\,1s
($<$v$>$=3000 \kms). There are no {\em Narrow Line Quasars},
i.e. there are no AGNs with quasar luminosities and FWHM(\hbeta)$<$2000
\kms. NLSy1s comprise 20--30\% of the AGN population at faint absolute
magnitudes (\MB $>-22$), irrespective of the selection method. In the
RASS sample we find $\Gamma$\,[0.1-2.4 keV] $<$3.3. The $\Gamma$ vs.
FWHM(\hbeta) distribution for Sy\,1 galaxies is consistent with
previous work.  For QSOs the spectral index also flattens with increasing 
FWHM(\hbeta), but they have larger linewidths than Seyfert 1s. 
\end{abstract}
\begin{keyword}
galaxies: active; galaxies: Seyfert; quasars: emission lines; X-rays: galaxies
\end{keyword}

\end{frontmatter}


\section{Introduction}
\vspace{-0.5cm}
Seyfert\,1 galaxies with small widths ($<$2000 \kms) of the permitted emission
lines coming from the broad line region (BLR) 
have been given a separate name {\em Narrow Line Seyfert\,1 galaxies 
(NLSy1)}, although they probably do not form a class physically separated 
from classical Sy\,1s. Close attention was drawn to them resulting from  the
discovery that steep soft X-ray spectra of AGNs are almost always
associated with NLSy1s \cite{Boller96}. It was also claimed that
X-ray selected samples contain considerably more NLSy1s than optically
selected ones \cite{Stephens89}. Since Sy\,1 galaxies are not the most 
luminous examples of AGNs, we searched for similar effects among quasars, their
high luminosity counterparts. We analyzed the Hamburg database of
optical AGN spectra to determine the fraction of NLSy1s in optically and 
X-ray selected samples, to search for {\em Narrow Line QSOs},
and to study the relation between soft X-ray spectral index and linewidth,
as a function of luminosity.

\section{The sample}
\vspace{-0.5cm}
The Hamburg database of optical AGN spectra consists of several data sets
obtained from follow-up spectroscopy of AGN candidates, identified from the
objective prism plates of the Hamburg Quasar Survey (HQS; \cite{Hagen95}).
The first set (hereafter: RASS)
contains spectra of X-ray selected AGNs discovered
by the ROSAT All-Sky Survey (RASS). 253 spectra were taken from \cite{Bade95}
and 189 spectra are yet unpublished. ROSAT X-ray selected AGNs have typical redshifts 
z=0.2 and optical brightnesses B=18--19 mag. 

\vspace{0.3cm}
\begin{figure}[htb]
\centerline{\psfig{figure=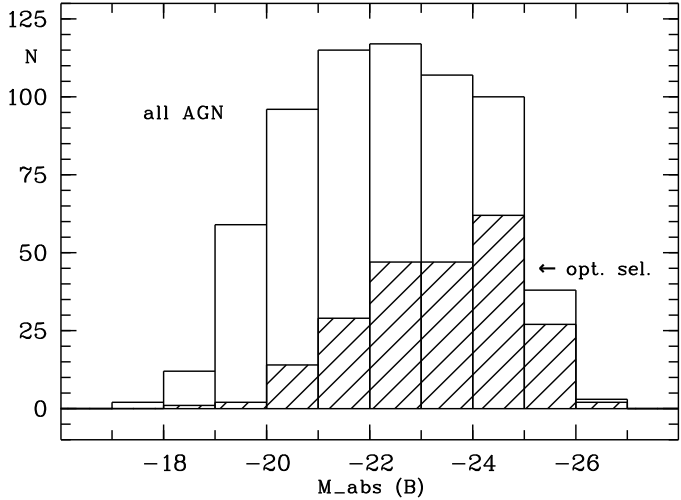,width=09cm,angle=-90}}
\caption{\label{mabs} 
The absolute magnitude distribution of the full sample (HQS + RASS).
The HQS sample is hatched.}
\end{figure}

The second set (hereafter: HQS) 
contains the 
spectra of AGNs optically selected from the HQS plates, and consists of 123
spectra from the lists of \cite{Engels98} and \cite{Hagen99}, and 112
spectra from the Hamburg/CfA Bright Quasar Survey \cite{Engels00}. The HQS AGNs
generally have B$<$18 mag. In both sets
we restricted the redshifts to z$<$0.7. 
The database is inhomogeneous, as the spectra differ in resolution and
signal-to-noise ratio. We singled out spectra with
sufficient signal-to-noise ratio to obtain reliable measures of 
linewidths using Gaussian fits. Because of the differing optical
limits of both sets, the HQS sample is biased towards quasars (defined
as AGNs with \MB $< -23$), and the
RASS sample is biased towards Seyfert galaxies (Figure \ref{mabs}).

\section{\hbeta\ linewidths and the search for ``Narrow Line Quasars''}
\begin{figure}[htb]
\begin{minipage}[t]{13.8cm}
 \begin{minipage}[t]{6.9cm}
  \begin{flushleft}
   \psfig{figure=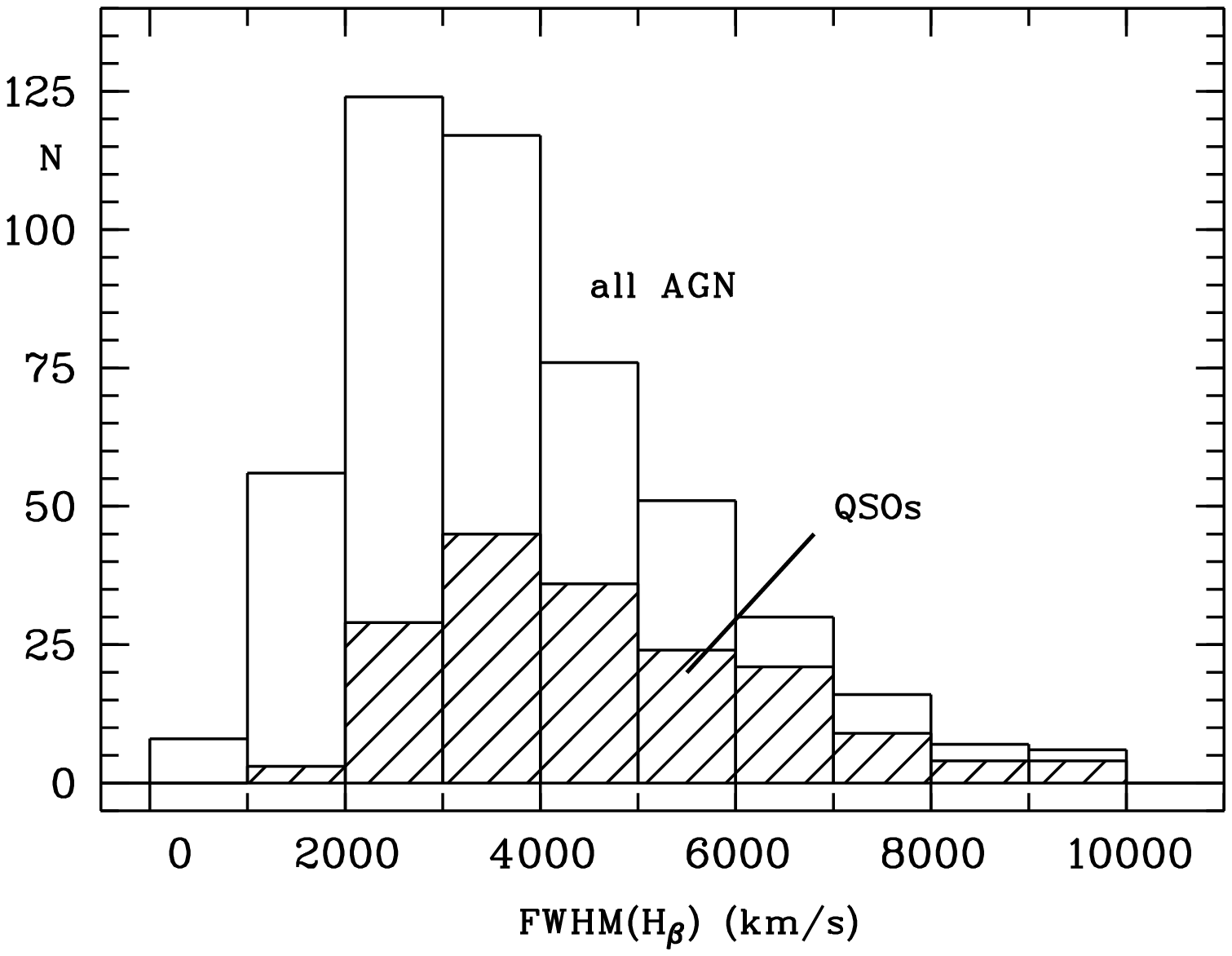,width=6.8cm,angle=-90}
  \end{flushleft}
 \end{minipage}
 \hfill
 \begin{minipage}[t]{6.9cm}
  \begin{flushright}
   \psfig{figure=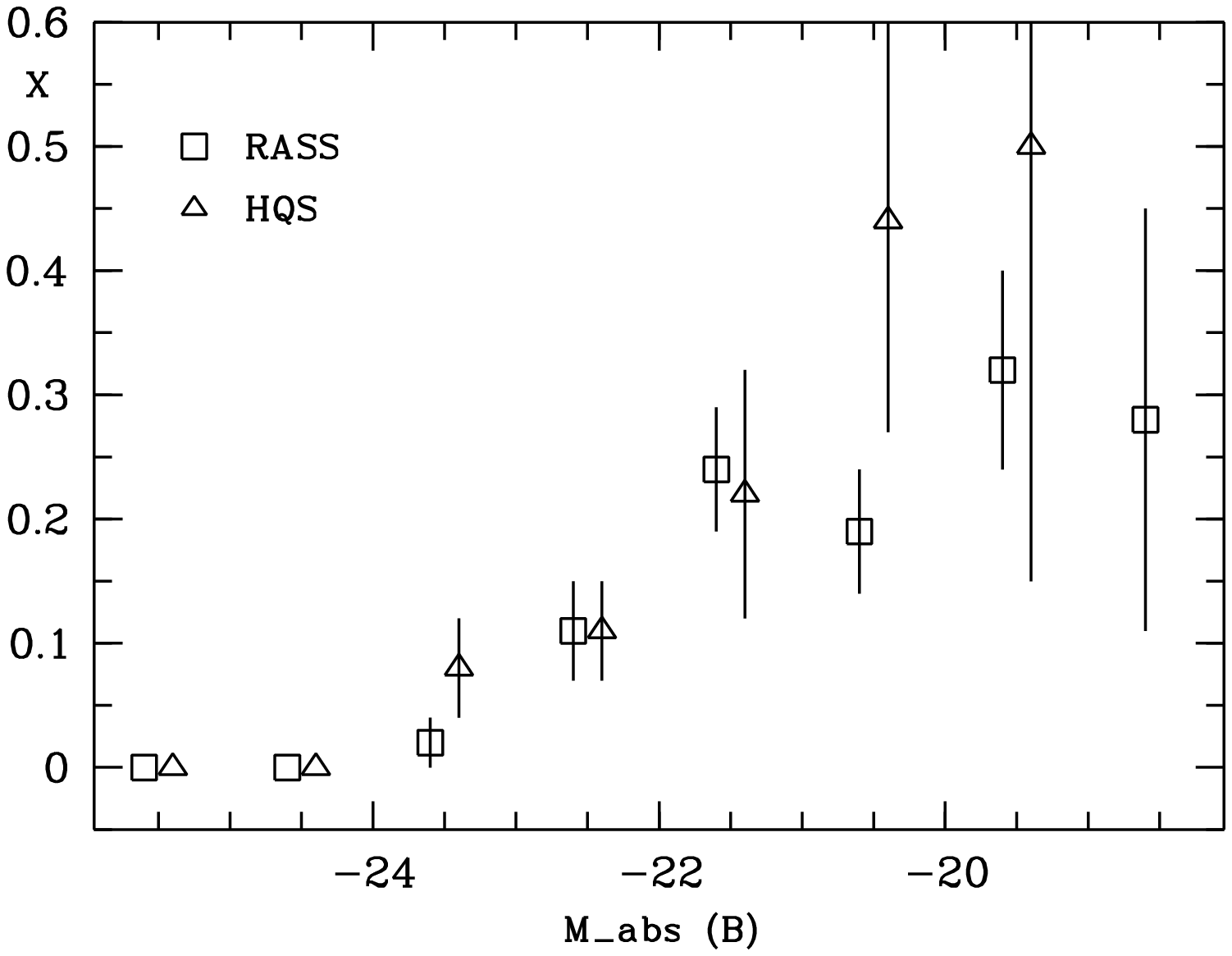,width=6.8cm,angle=-90}
  \end{flushright}
 \end{minipage}
\end{minipage}
\caption[]{ \label{hbeta} Left:
\hbeta\ linewidth distribution of the full sample (HQS + RASS).
The distribution of the QSOs is hatched. Right: 
Fraction $\chi$ of Narrow Line AGNs per luminosity bin as function of absolute
optical magnitude. Symbols are offset by 0.1 around the bin center
for clarity.} 
\end{figure}

\vspace{-0.5cm}
The \hbeta\ linewidth distribution is markedly different for the 
Seyfert galaxies and the quasars (Figure \ref{hbeta}a). 
The latter group has a median linewidth
of $<$v$>$ = 4300 \kms\ compared to 3000 \kms\ for the Seyferts.
This difference does not depend on the selection method. Both
distributions have a long tail extending up to 10000 \kms. 

%
%
\vspace{0.3cm}

The handful of quasars with FWHM(\hbeta)$<$2000 \kms\ only marginally
surpass the luminosity cut-off, which was set arbitrarily. In
Figure \ref{hbeta}b we plot the fraction $\chi$ of
Narrow Line AGNs as function of absolute magnitude. In both samples
none are found at luminosities \MB $< -23.5$, indicating that broad-line 
regions with very small velocities exist only among low luminosity AGNs: There 
are no {\em Narrow Line QSOs}. The non-existent NLQSOs should not be confused
with {\em Type 2 QSOs}, which are the high-luminosity counterparts of
Seyfert 2s, and have not been found either. The fraction of NLSy1s rises
to 20--30\% for \MB $> -22$, irrespective of the selection method.

\vspace{-0.5cm}
\section{\hbeta\ linewidths and X-ray spectral index}
\vspace{-0.5cm}

\begin{figure}[htb]
\centerline{\psfig{figure=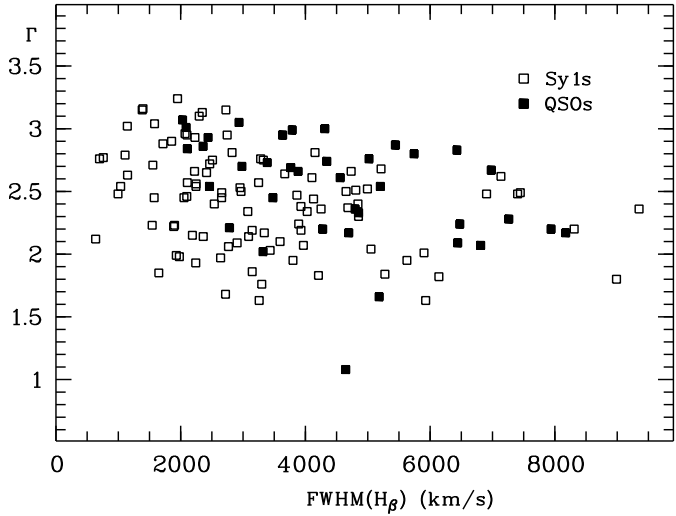,width=09cm,angle=-90}}
\caption{\label{gamma} 
ROSAT X-ray spectral slope $\Gamma$\,[0.1-2.4 keV] vs. \hbeta\ linewidth for
RASS AGNs.}
\end{figure}

X-ray spectral photon indices $\Gamma$ were calculated from the
hardness ratios as described by \cite{Bade95}.  Figure \ref{gamma}
plots $\Gamma$ vs.  FWHM(\hbeta). Seyfert galaxies occupy the same
area as in \cite{Boller96}, except that the sample contains no {\em
ultrasoft} X-ray AGNs ($\Gamma > 3.3$). Both Seyfert galaxies and
quasars show a modest anti-correlation between $\Gamma$ and \hbeta\
linewidth. This is consistent with the model of Wandel \& Boller
\cite{Wandel97,Wandel98}, who attribute this relation to an
increased ionizing flux associated with the steeper X-ray spectral slopes.
For a given black hole mass this shifts the BLR radius outwards leading 
to smaller velocities.

The \hbeta\ linewidths are thought to reflect the velocity dispersion in the
BLR. We found that this 
dispersion increases with AGN luminosity, as does the BLR radius
\cite{Kaspi96}. This suggests that the central black hole masses must
increase as well, if the cloud motions are virialized. Narrow Line AGNs
are common among low luminosity AGNs (Seyferts), and their X-ray
properties in general do not differ from other AGNs. The larger number of
NLSy1s detected in X-ray surveys is due to the higher selection efficiency
of Seyfert 1s in general, compared to optical surveys. There appear to be no
physical differences between X-ray and optically selected Narrow Line
AGNs.

We have not found objects with extreme $\Gamma$'s ($>3.3$), which were
so obvious in the sample of \cite{Boller96}.  It seems that these
{\em ultrasoft} AGNs are very rare, and that they are only picked-up in 
surveys with high efficiency in the soft X-ray band.
They may not be restricted to low luminosities, as
at least one example in known eg. RX~J0947.0+4721 with a QSO luminosity 
\cite{Molthagen98}.  RX~J0947.0+4721 was discovered in a
pointed ROSAT observation, and the RASS might not have been sensitive
enough to detect more of them. The rarity of {\em ultrasoft} AGNs
may be a consequence of the transient nature of the {\em ultrasoft}
state, which may be short compared to the timescales that the nuclei are
active.

{\bf Acknowledgements:} 
We like to thank A. Dobrzycki, M. Elvis, H.-J. Hagen, J. Hu, C. Ledoux,
F. Tesch, A. Ugryumov, D. Valls-Gabaud, and J. Wei for providing
 optical spectra in advance of publication.







\end{document}